\documentclass[12pt,a4paper]{article}

\usepackage[T1]{fontenc}
\usepackage[utf8]{inputenc}
\usepackage{amsmath}
\usepackage{amssymb}
\usepackage{graphicx}
\usepackage{float}
\usepackage{color}
\usepackage[margin=2cm]{geometry}
\usepackage{appendix}
\usepackage{multirow}

\usepackage[unicode]{hyperref}
\usepackage{booktabs}
\usepackage{comment}
\usepackage{bm}

\usepackage[style=authoryear,maxcitenames=1,natbib=true,uniquelist=false,sorting=nyt,backend=bibtex]{biblatex} 
\usepackage{wrapfig} 
\usepackage{subcaption}
\hypersetup{colorlinks,urlcolor=black,citecolor=black,linkcolor=black,filecolor=black}
\usepackage{enumitem}
\usepackage{siunitx}

\usepackage{orcidlink}
\newcommand{\orcidGB}{\orcidlink{0000-0003-2849-0120}} 
\newcommand{\orcidGGB}{\orcidlink{0000-0001-9223-6480}} 
\newcommand{\orcidGP}{\orcidlink{0000-0001-5038-678X}} 
\newcommand{\orcidZJ}{\orcidlink{0009-0001-2300-3605}} 

\title{Comparison of Proton, Helium-Ion, and Carbon-Ion CT Imaging for Hadron Radiotherapy Using Monte Carlo Simulations}

\author{
Zsófia Jólesz$^{1,2,*}$\orcidZJ, Gábor Bíró$^{1}$\orcidGB, Gábor Papp$^{2}$\orcidGP and Gergely Gábor Barnaföldi$^{1}$\orcidGGB\\[1em]
\parbox{0.9\textwidth}{\centering\small
$^{1}$HUN-REN Wigner Research Centre for Physics,\\
29--33 Konkoly--Thege Mikl\'os \'ut, H-1121 Budapest, Hungary\\[0.3em]
$^{2}$E\"otv\"os Lor\'and University, Institute of Physics and Astronomy,\\
1/A P\'azm\'any P\'eter S\'et\'any, H-1117 Budapest, Hungary\\[0.6em]
$^{*}$Author to whom any correspondence should be addressed: \texttt{jolesz.zsofia@wigner.hun-ren.hu}
}
}

\date{}

\begin{document}

\maketitle

\begin{abstract}

\noindent \textbf{Objective:} To perform a comprehensive comparative analysis of proton, helium-ion, and carbon-ion computed tomography (CT) as direct imaging modalities for hadron therapy treatment planning, focusing on Relative Stopping Power (RSP) reconstruction accuracy and patient radiation dose.

\noindent \textbf{Approach:} High-fidelity Monte Carlo simulations were conducted using the GATE/Geant4 platform to model a standard CTP404 phantom. RSP maps were reconstructed using two iterative algorithms: the Simultaneous Algebraic Reconstruction Technique (SART) and the Richardson-Lucy (RL) deconvolution. Imaging performance was evaluated by comparing reconstructed RSP values against ground truth data for various tissue-equivalent inserts, while integral doses were estimated for a human head geometry using a Shepp-Logan-like phantom.

\noindent \textbf{Main results:} All investigated particle modalities demonstrated a significant dose reduction compared to conventional X-ray CT protocols (which are approximately 40 mGy). The estimated imaging doses were $0.5$ mGy for protons, $1.3$ mGy for helium ions, and $7.5$ mGy for carbon ions. 
In terms of accuracy, carbon-ion imaging achieved promising accuracy for soft-tissue materials (mean absolute error $<0.5\%$). Helium ions offered a balanced performance with sub-$1\%$ errors for most materials and a dose burden significantly lower than carbon ions. Protons exhibited the widest range of RSP deviations.

\noindent \textbf{Significance:} Direct particle imaging eliminates the systematic uncertainties inherent in photon-to-hadron conversion, which typically introduce RSP deviations of 1.6\%–3.5\% in standard single-energy X-ray CT. While carbon ions provide better RSP reconstruction precision essential for complex treatment plans in heterogeneous anatomy, helium and proton imaging offer exceptional dose sparing, making them particularly advantageous for pediatric patients and frequent adaptive replanning scenarios.
\end{abstract}

\vspace{0.5pc}

\section{Introduction} \label{sec:introduction}
\subsection{Hadron Therapy}
Since cancer remains one of the leading causes of mortality worldwide, the optimization of cancer treatment continues to be an area of intense scientific investigation and has undergone substantial development over recent decades. Among the principal modalities of cancer therapy is radiotherapy, the overarching objective of which is to irradiate malignant tissue as accurately as possible while minimizing radiation exposure to surrounding healthy organs. A rapidly advancing branch of radiotherapy involves the use of hadrons (e.g.\ protons or heavier ions such as helium or carbon), which exhibit particularly advantageous physical properties for oncological applications due to their characteristic energy deposition profile, described by the Bragg curve (\cite{bragg1904lxxiv, Dickmann_2019}). This phenomenon enables highly localized dose delivery, allowing the Bragg peak---the point of maximal energy deposition---to be positioned within the tumor volume while substantially reducing the dose imparted to adjacent healthy tissues.

\begin{figure}[htbp]
\centering
\includegraphics[width=0.95\textwidth]{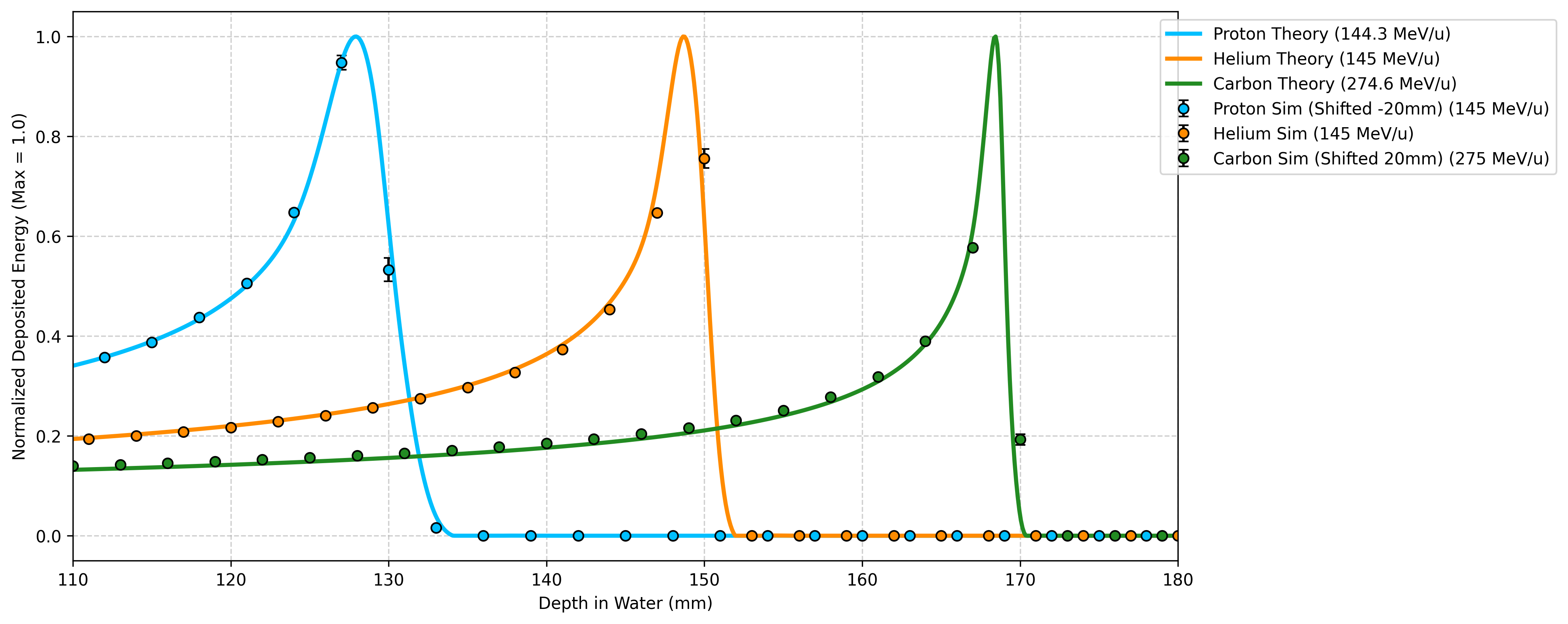}
\caption{Simulated and fitted Bragg peaks for protons, helium ions and carbon ions. The fit was obtained by integrating over the energies in the Bragg curve (\cite{bragg1904lxxiv}) sampled from a normal distribution around the most probable energy, accounting for the spread. For the better visibility, the proton and the carbon curves are shifted with -20 and 20 mm, respectively.}
\label{fig:bragg}
\end{figure}

The Bragg peak is even narrower for helium and carbon ions than for protons (Figure~\ref{fig:bragg}), suggesting that helium- and carbon-ion therapy may, in principle, provide superior tumor conformity and reduced side effects (\cite{byun2021physical}, \cite{wickert2022radiotherapy}). As the clinical implementation of these ion species has become feasible, comparative evaluation of different hadron-therapy modalities has emerged as a major research focus. \cite{wickert2022radiotherapy} reported that the use of helium ions instead of conventional photon Intensity Modulated Radiation Therapy (IMRT) resulted in a significant reduction in the integral dose absorbed by healthy brain tissue ($-45.3\%\pm15.1\%$). Even when compared to actively scanned protons, helium ions achieved a further relative dose reduction of $-27.2\%\pm14.4\%$ for critical neuronal structures.

\cite{huang2022comparison} assessed acute dermatitis as an adverse effect, finding that among patients treated with proton therapy, 16 of 18 developed acute grade I/II dermatitis, whereas only 7 of 20 patients treated with Carbon Ion Radiotherapy (CIRT) experienced acute grade I dermatitis.

While heavier ions offer superior tumor control, they also carry higher risks for surrounding healthy tissues due to nuclear fragmentation and higher Linear Energy Transfer (LET). Literature indicates that helium ions deposit roughly hundred times as much physical dose as that of the proton, while in the case of carbon ions, this can be as much as a thousand times greater
(\cite{payandeh2025comparative}). However, the actual cumulative dose of an imaging scan is highly dependent on the total number of primary particles required to reach a stable reconstruction. Therefore, a detailed calculation of the cumulative physical imaging doses (in mGy) is provided in Section \ref{sec:discussion} of this study. 
Because the potential damage to healthy tissues is higher for helium and carbon ions, ensuring extreme precision in dose delivery is highly important, therefore highly accurate Relative Stopping Power mapping is even more critical for heavy ion therapy than for proton therapy.

\subsection{Imaging for Hadron Therapies}
In current clinical practice, treatment planning for hadron therapy is predominantly based on X-ray computed tomography (CT) imaging (\cite{kaiser2019proton}). While X-ray CT provides electron density information of the irradiated tissues, this information must be converted into relative stopping power (RSP) maps in order to perform dose calculations for charged-particle therapy~\cite{10.1088/0031-9155/60/19/7585}. This conversion relies on empirical calibration curves and introduces systematic uncertainties due to the fundamentally different interaction mechanisms of photons and hadrons with matter (\cite{ordonez2019fast, Dickmann_2019, 10.1002/mp.15644}). \cite{hansen2015simulation} demonstrated that single-energy X-ray CT leads to an RSP uncertainty of approximately 2.7\%, which can be reduced to 0.5--0.9\% using dual-energy CT; however, the latter approach is technically more complex and costly and while they can reduce them, image artifacts can still be present with this type of imaging (\cite{borges2023pros}). Depending on the anatomical site and the required penetration depth, this relative error can propagate into several millimeters of absolute range uncertainties in clinical scenarios (\cite{paganetti2012range}) and may compromise accurate dose delivery, particularly in regions with heterogeneous tissue composition such as the brain, lungs, or head-and-neck area. Such range uncertainties necessitate the use of additional safety margins, which in turn limit the full exploitation of the steep distal dose fall-off characteristic of hadron beams.

A promising strategy to mitigate these uncertainties is to employ the same type of particle for imaging as for treatment, thereby eliminating the need for photon-to-hadron conversion. In the case of proton therapy, this concept has led to the development of proton computed tomography (pCT), a novel imaging modality that directly reconstructs the RSP distribution of the patient. Recent studies have demonstrated the feasibility of pCT and reported substantial reductions in range uncertainty compared to conventional X-ray CT-based treatment planning, highlighting its strong potential for improving dose accuracy in proton therapy.

While direct particle imaging is highly promising, its clinical applicability relies on the development of specialized detection systems capable of tracking single ions and measuring their residual energies. Several state-of-the-art prototype systems have already been developed for this purpose, including the Phase II proton CT scanner at Loma Linda University/UCSC (\cite{bashkirov2016development}), the INFN pCT system (\cite{scaringella2023infn}), and the highly granular digital tracking calorimeter developed by the Bergen pCT Collaboration, which utilizes the MAPS (Monolithic Active Pixel Sensor) technology also applied in the ITS (Inner Tracking System) of the ALICE experiment at CERN (\cite{BergenpCTStatusReport}). For transmission imaging, these systems require initial beam energies sufficiently high to traverse the patient entirely and place the Bragg peak inside the detector. For adult cranial imaging (meaning typically 15--20 cm water-equivalent thickness), this demands clinical beam energies of approximately 200--230 MeV/u for protons and helium ions. For carbon ions, due to their higher energy loss rate, an initial energy of up to 430 MeV/u is required to achieve a similar 30 cm water-equivalent range. Such energies are available at dedicated clinical facilities, such as the Heidelberg Ion Therapy Center (HIT), establishing the physical feasibility of multi-ion computed tomography (\cite{haberer2004heidelberg}).

As radiotherapy with heavier ions such as helium and carbon gains increasing clinical relevance due to their favorable physical and biological properties, the question naturally arises whether particle imaging with these ions could offer similar or even superior benefits for treatment planning. However, systematic investigations comparing proton-, helium-, and carbon-ion imaging are still limited. Therefore, the objective of this work is to perform a comprehensive comparison of imaging with protons, helium ions, and carbon ions, with particular emphasis on their cumulated dose deposition, their respective imaging performance, achievable range resolution, and potential implications for clinical treatment planning. By evaluating the advantages and limitations of each particle species, this study aims to contribute to the optimization of future image-guided hadron therapy.

\section{Methods} \label{sec:methods}
\subsection{Simulations}
\begin{figure}[htbp]
    \centering
    \begin{subfigure}[b]{0.48\textwidth}
        \centering
        \includegraphics[width=\textwidth]{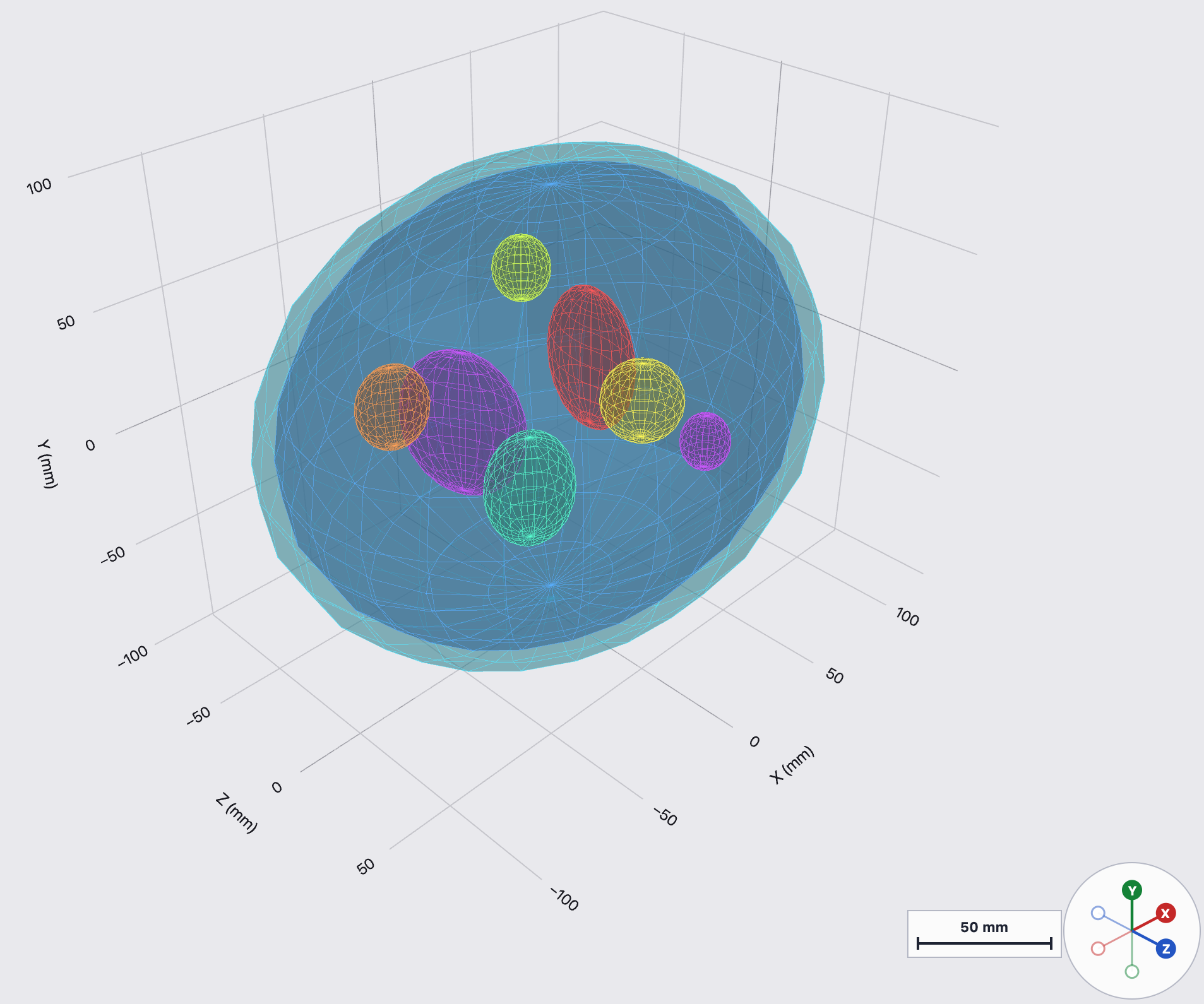}
        \caption{3D rendering}
        \label{fig:shepp_logan_3d}
    \end{subfigure}
    \hfill
    \begin{subfigure}[b]{0.48\textwidth}
        \centering
        \includegraphics[width=\textwidth]{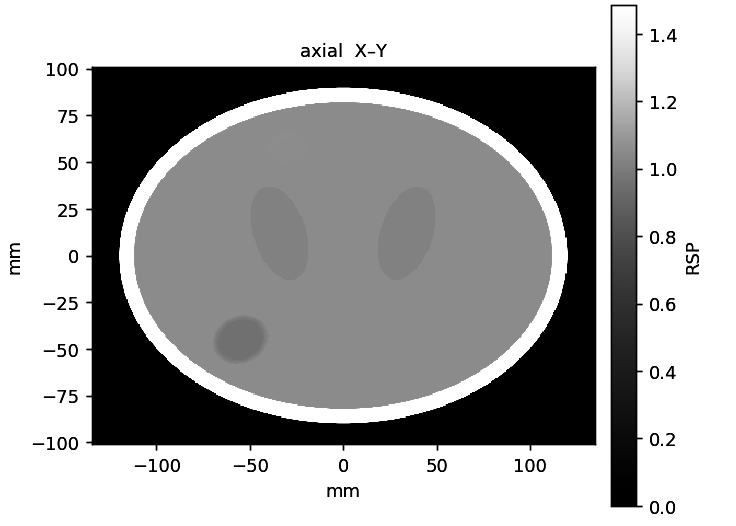}
        \caption{Axial cross-section}
        \label{fig:shepp_logan_cross_sec}
    \end{subfigure}
    \caption{Visual representation of the 3D Shepp-Logan-like head phantom utilized for the absolute dose deposition estimation. (a) 3D rendering of the external hull and internal tissue-equivalent ellipsoidal structures. (b) Axial cross-sectional slice (X-Y plane).}
    \label{fig:shepp_logan_phantom}
\end{figure}

For this work, the data taking was done by Monte Carlo simulations, using Geant4 (version 11.4.2)~\citep{Agostinelli2003, Allison2006} and GATE 10 toolkit~\citep{sarrut2026gate, Jan2004, Jan2011} to model the interactions of charged particle beams with a phantom. Two different phantoms were used in this study taking two main goals into account: to test the RSP reconstruction accuracy and measure the dose deposition of the particles. The CTP404 phantom (\cite{CTP404}) is designed to evaluate the accuracy of material density reconstruction in homogeneous regions. It is a 400 mm long disk with a diameter of 150 mm and contains eight cylindrical rods with a diameter of 12.2 mm, each composed of a different material. The other phantom utilized in this study was a Shepp-Logan-like phantom (\cite{shepp1974fourier}). This phantom contains 3-dimensional ellipsoids that have the same RSP values as the various tissues in a human head (such as skull bone, brain, muscle, blood, cartilage and fat). The 3-dimensional rendering and the cross-sectional view of the phantom can be seen on Figure \ref{fig:shepp_logan_phantom}. While it fairly simplifies a human head, the realistic materials and the dimensions of the phantom make it suitable for a good estimation of the dose deposition.

The simulations utilized the QGSP\_BIC\_EMY physics list to govern hadronic and electromagnetic processes, with the ionization potential of water manually adjusted to 75 eV. The radiation source was modeled as a Pencil Beam Scanning (PBS) system originating at $Z = -500$ mm. For all cases, the beam was monoenergetic ($\sigma_E = 0.0$ MeV) and characterized by a Gaussian spatial profile with $\sigma_x = \sigma_y = 3$ mm, an angular divergence of 2.8 mrad, and an emittance of 3.0 mm$\cdot$mrad, mimicking the realistic beam parameters of the CNAO facility \citep{giordanengo2015cnao}. In a clinical practice, transmission imaging is ensure by selecting sufficiently high initial beam energies to traverse the maximum thickness of the patient, with the residual energy fully absorbed by a deep range telescope or calorimeter. To investigate different particle species, three distinct beam configurations were simulated, the protons and the alpha particles maintaining a constant specific energy of 230 MeV/u, while the Carbon-12 ions had 430 MeV/u initial energy. The higher initial energy assigned to the carbon beam compensates for its significantly faster rate of energy loss, ensuring that the Bragg peaks for all three particle species successfully penetrate the phantom and are positioned roughly at the same depth in the downstream detector region. While carbon ions provide highly advantageous multiple Coulomb scattering properties, their transport is heavily influenced by nuclear fragmentation. In our simulations, a filter is applied, ensuring that only the intact, primary ions are used for the reconstruction later. As a result of this, only $\sim60\%$ of the primary carbon ions traversed the phantom without being fragmented. In a clinical scenario, accurately estimating the physical range the particle travels before stopping at the Bragg peak requires sufficient hardware to reject these secondary fragments. Highly granular detector concepts, such as the digital tracking calorimeter developed by the Bergen pCT Collaboration \citep{DesignPixelRangeTelescope, BergenpCTStatusReport}, are capable of topologically distinguishing intact primaries from lighter secondary fragments. 

Unlike therapeutic PBS treatment plans, which actively modulate beam energy for volumetric dose coverage, the transmission imaging sequence utilizes a constant initial beam energy for all scanning positions. The simulated energies were therefore maintained at the previously mentioned values in order to mimic a real-life imaging scenario.

To record the position, direction, and energy of the exiting particles, a \texttt{PhaseSpaceActor} was placed at a longitudinal position of $Z = 400$ mm within a $500 \times 500 \times 100$ mm$^3$ scoring volume. While this transmission imaging setup is fundamentally motivated by the development of novel tracking calorimeters, such as the Bergen pCT Collaboration system. However, this study is based on an ideal detector framework, not considering the detector effects during imaging. In clinical practice, realistic hardware constraints, such as electronic noise, finite spatial and energy resolution, and digitization effects are known to degrade image quality. For instance, recent study by \cite{bíró2026protoncomputedtomographyimage} demonstrated that such detector effects can increase the relative RSP reconstruction error from $0.7\%$ (for an ideal detector) up to $2.7\%$ (for silicon strip configurations), while substantially reducing spatial resolution (MTF$_{10\%}$). However, simulating these hardware-specific uncertainties falls outside the scope of the current work, which focuses exclusively on isolating and comparing the fundamental physical imaging capabilities of the respective ion species.

\subsection{Dose Calculation}

To accurately evaluate the physical radiation dose delivered to a patient during a full tomographic scan, a 3-dimensional Shepp-Logan-like head phantom was utilized directly within the OpenGATE framework. The phantom features an elliptical outer skull (composed of cortical bone material with an RSP $\approx1.6$) with in-plane semi-axes of 120 mm and 90 mm, and a height half-extent of 95 mm. The internal cavity contains a brain-equivalent soft tissue background (RSP $\approx1.0-1.1$), embedded with several anatomically representative, 3-dimensional ellipsoids. These ellipsoids simulate lateral ventricles (water/cerebrospinal fluid), cartilage, deep bone nodules, adipose tissue, muscle regions and blood. The total mass of the simulated head phantom was calculated by the Geant4 geometry engine and is approximately 5 kg.

In order to capture the absolute physical dose, a \texttt{DoseActor} was attached directly to the phantom's local frame. The actor makes it possible to record the exact total energy deposited ($E_{dep}$) inside the phantom volume by every interacting particle. This approach accounts for the dose contributions from primary ionization, delta rays, and all secondary nuclear fragmentation products, which is a critical consideration for heavy-ion transport.

The scanning of the phantom consisted of 360 rotations (every degree), 132 translations along the phantom hull's cross-section, and 32 translations along the vertical axis, resulting in $360\times132\times32\times300$ primary particles traversing the phantom. The cumulative energy deposited by these primaries and the previously mentioned secondary products was aggregated. The mean physical absorbed dose, expressed in Gray was then explicitly calculated by converting the total deposited energy into Joules and dividing it by the exact mass of the phantom:
\begin{equation}
D_{mean} = \frac{\sum E_{dep} \times (1.602 \times 10^{-13} \text{ J/MeV})}{m_{phantom}}
\end{equation}
The method was repeated for all beam types with the previously described initial energy values (230 MeV/u for protons and helium ions and 430 MeV/u for carbon ions).

\subsection{Calculation of RSP Values}
To establish the baseline Ground Truth RSP values for the specific materials, individual per-material GATE10 simulations were performed. The geometry consisted of $60\times60\times60$ cm block phantoms assigned to each distinct material found in the CTP404 phantom and water, within $1.5\times1.5\times1.5$ m of vacuum. The primary particle beams were simulated interacting with these blocks, and the 1-dimensional depth-dose curves were scored utilizing a \texttt{DoseActor} with a longitudinal spatial resolution of 0.01 mm along the beam axis. The physical range of the particles in each material was defined as the physical distance the particle travels before reaching its peak energy deposition, which was determined by extracting the spatial coordinate of the Bragg peak from the depth-dose distributions. The baseline RSP for each non-water material was subsequently calculated by taking the ratio of the simulated physical range in water to the physical range in the respective target material. To ensure the physical validity of this setup, the simulated physical ranges and resulting RSP values were cross-validated against theoretical values provided by the NIST PSTAR database \citep{berger1992estar} and specifically helium simulations were compared with RSP values from \cite{piersimoni2018helium}.
Using 200 MeV/u initial energy, the relative difference between the values from \cite{piersimoni2018helium} and the simulated RSP values was within 2\%, while it remained within 1\% when compared to the NIST PSTAR values.

Following successful validation, the simulations were extended to protons and carbon ions, using the initial energies that were later used for the imaging simulations (230 MeV/u for protons and helium ions, and 430 MeV/u for carbon ions). It is important to note that the negligible difference in the RSP values between the different particle species is due to the high-energy transmission regime. Initial kinetic energies required for transmission imaging ensure that the ions remain fully stripped of electrons across the majority of the phantom. Consequently, the $Z^2$ term in the Bethe-Bloch formulation cancels out during normalization to water, resulting in similar RSP values across the distinct particle species. This phenomenon is also demonstrated in \cite{hurley2012water}.

\subsection{Image Reconstruction}
To reconstruct the RSP distributions, the problem is formulated as a large sparse linear system of equations ($Ax=b$), where $b$ represents the measured WEPL of the individual primary particles, $x$ is the voxelized RSP map, and the system matrix $A$ contains the geometric intersection lengths of each particle's path with the reconstruction grid voxels. The entry and outgoing coordinates are calculated using a Most Likely Path (MLP) formalism by \cite{schulte2008maximum, krah2018comprehensive}. Inside the phantom, the different multiple Coulomb scattering characteristics of the distinct ion species are accounted for during the construction of the system matrix by applying a third-order spline approximation (\cite{williams2004most}).

In order to demonstrate that the reconstruction quality is robust and not biased by the choice of the numerical solver, two distinct iterative algorithms were utilized for reconstruction: the algorithm based on the Richardson-Lucy (RL) method and the Simultaneous Algebraic Reconstruction Technique (SART) \citep{bíró2026protoncomputedtomographyimage, andersen1984simultaneous}. The solvers utilize the trajectories obtained from the previously mentioned MLP calculation. The RL algorithm acts as a Maximum Likelihood Expactation Maximization (MLEM) solver, employing a multiplicative update rule defined as $x_j^{(k+1)}=x_j^{(k)}\cdot C_j$, where $C_j$ is a ratio-based correction factor derived from the backprojected data. In contrast, SART utilizes an additive update strategy, taking the form $x_j^{(k+1)}=x_j^{(k)}+\lambda \cdot E_j$, where $\lambda$ is a relaxation parameter and $E_j$ represents the normalized projection error. Both solvers iteratively minimize the projection data misfit.

To ensure algorithmic reproducibility and maintain identical statistical conditions across all imaging modalities, the reconstruction parameters were uniform. The total number of utilized primary particle tracks was fixed at $25\times10^6$ for all three particle species. This specific threshold was constrained by the carbon-ion simulations, representing the total number of primary carbon ions that successfully traversed the phantom without undergoing nuclear fragmentation or large angle scattering. Both the RL and SART solvers were implemented utilizing a block-iterative schema, executing a total of 1250 block. For each block, a subset of 20,000 particle tracks was processed with $100$ iterations to update the RSP map. The reconstruction was uniformly terminated upon the complete processing of the 25 million tracks, which served as a fixed stopping criterion. To evaluate the raw baseline imaging capabilities of the distinct particle species, no additional spatial regularization or noise-suppression was applied during the iterative updates. Instead, final image regularization was achieved by applying a Ruppert-Polyak averaging process (\cite{polyak1990new}) on the 1250 sequential volumetric RSP maps, effectively suppressing statistical noise.

There are other various reconstruction strategies that have been developed for this purpose, such as the Filtered Back Projection (FBP) along most likely paths \citep{rit2013filtered} or block-iterative methods \citep{penfold2010block}. However, evaluating or optimizing reconstruction algorithms is beyond the scope of the present study. While several previous studies have investigated the comparative imaging performance of various ion species, our objective here is to strictly assess the intrinsic imaging capabilities of different particle species under identical algorithmic conditions. A detailed comparison of our findings with these existing literature benchmarks is presented in Section 4.2.

\subsection{Evaluation}
The quantitative evaluation of the reconstructed RSP distributions was performed using the standard CTP404 phantom. Ground truth (GT) RSP values for the phantom's material inserts were derived directly from the GATE10 simulations, as described in Section 2.3.

To maintain consistency, the evaluation of reconstruction accuracy followed the exact Region of Interest (ROI) based methodology described in \cite{bíró2026protoncomputedtomographyimage}. By comparing the mean reconstructed RSP values within these predefined ROIs to the GT values, we quantified the material-dependent reconstruction fidelity for protons, helium ions, and carbon ions. 

Finally, by synthesizing this RSP accuracy analysis with the previously described dose estimations, we achieved a comprehensive and unbiased comparative evaluation of the imaging performance and dose efficiency of the three particle modalities under identical reconstruction conditions.

\section{Results \label{sec:results}} 
\subsection{Image Reconstruction Accuracy}

\begin{figure}[htbp]
    \centering
    \includegraphics[width=0.8\textwidth]{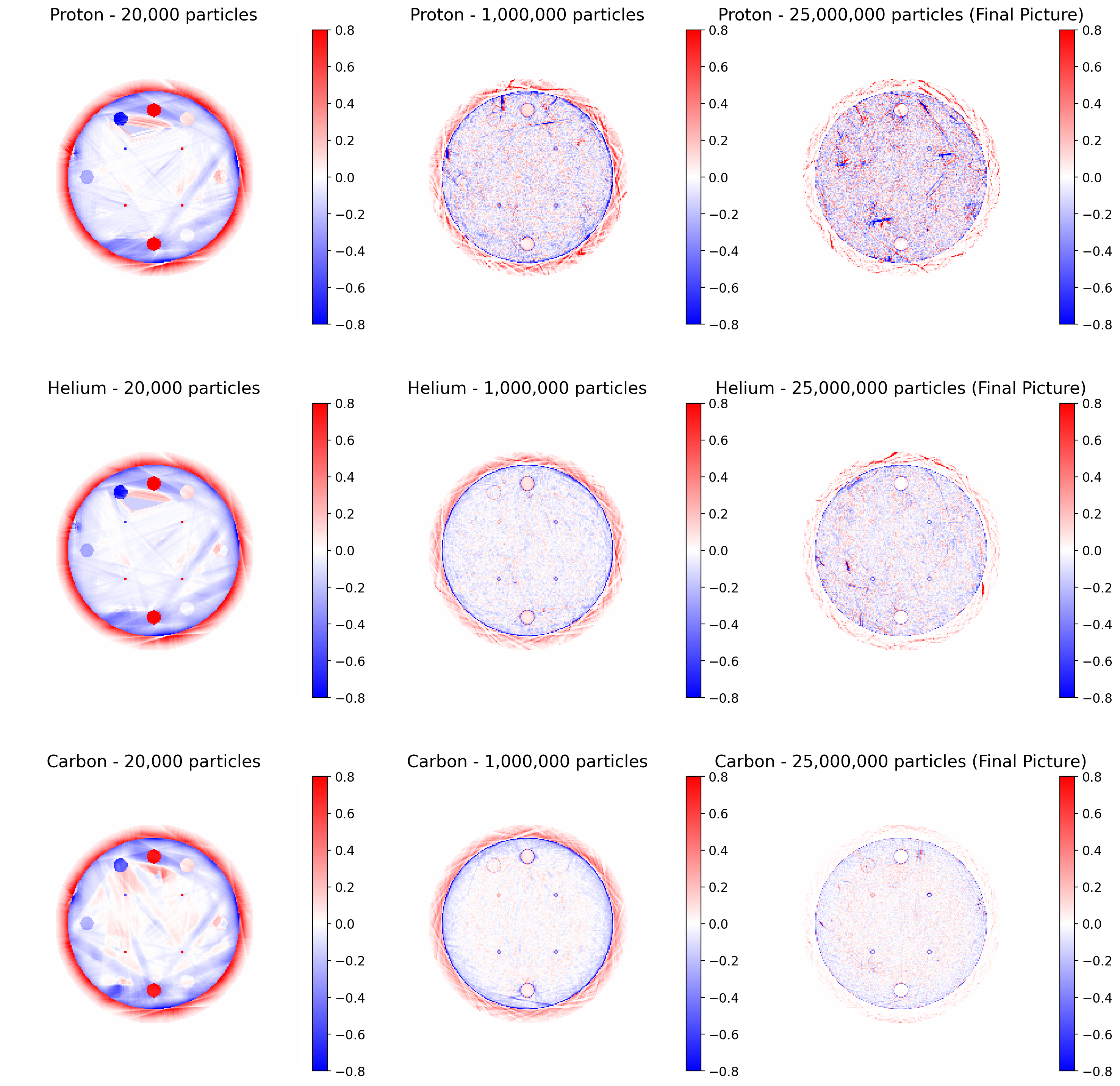}
    \caption{Absolute difference between the ground truth and the reconstruction values per pixel ($\Delta=\text{GT}-\text{Rec.}$) using different particle numbers.
    }
    \label{fig:diff_heatmap}
\end{figure}

To evaluate the overall reconstruction accuracy and the evolution of the reconstructed Relative Stopping Power (RSP) distributions, the difference between the ground truth (GT) and the reconstructed maps was first analyzed. Figure \ref{fig:diff_heatmap} presents these difference heatmaps for the entire CTP404 phantom across protons, helium ions, and carbon ions at varying particle statistics. The heatmaps visually demonstrate the progressive reduction in background noise and reconstruction artifacts as the number of processed particles increases towards the final baseline.

\begin{figure}[htbp]
    \centering
    \includegraphics[width=\textwidth]{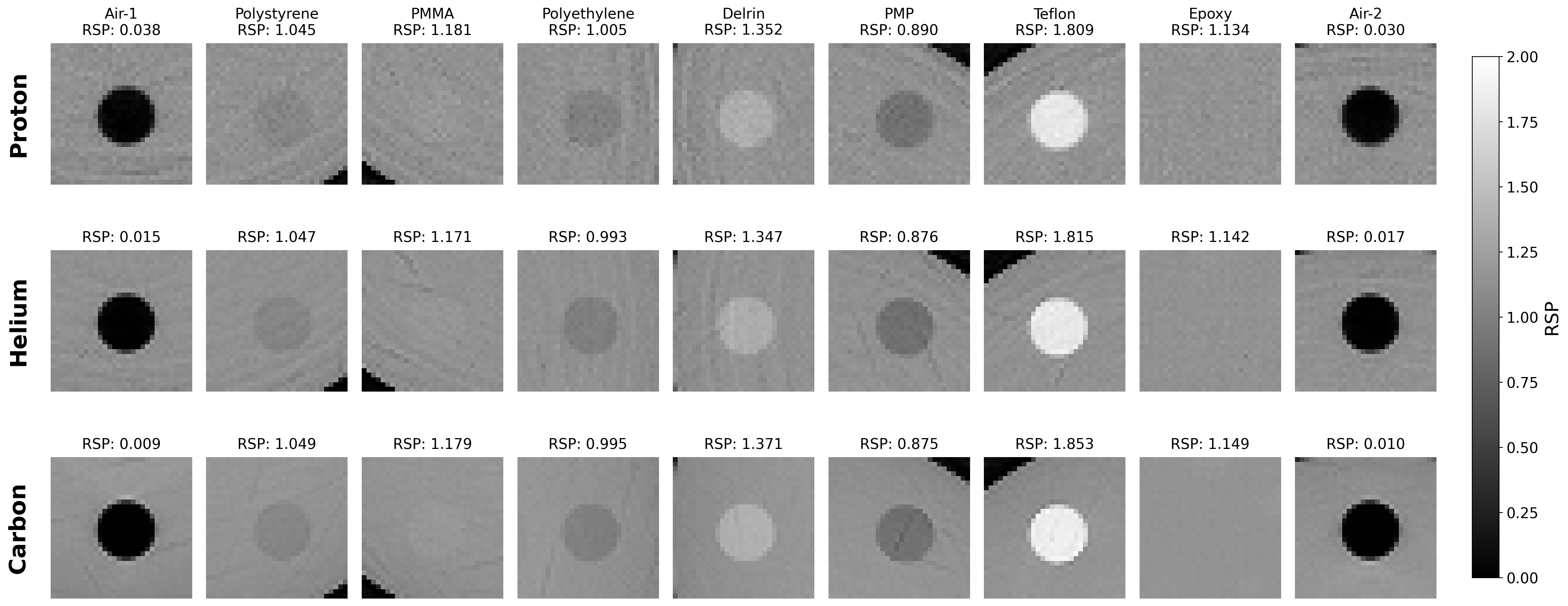}
    \caption{The reconstructed CTP404 inserts using protons, helium ions and carbon ions.}
    \label{fig:CTP404_rec}
\end{figure}

To more closely examine the material-specific reconstruction fidelity, the visual analysis was narrowed down to the individual material inserts. Figure \ref{fig:CTP404_rec} displays these zoomed-in Regions of Interest (ROIs) for the final reconstructed images of each particle modality, demonstrating which segments were utilized for the data extraction.

\begin{figure}[htbp]
    \centering
    \includegraphics[width=0.8\textwidth]{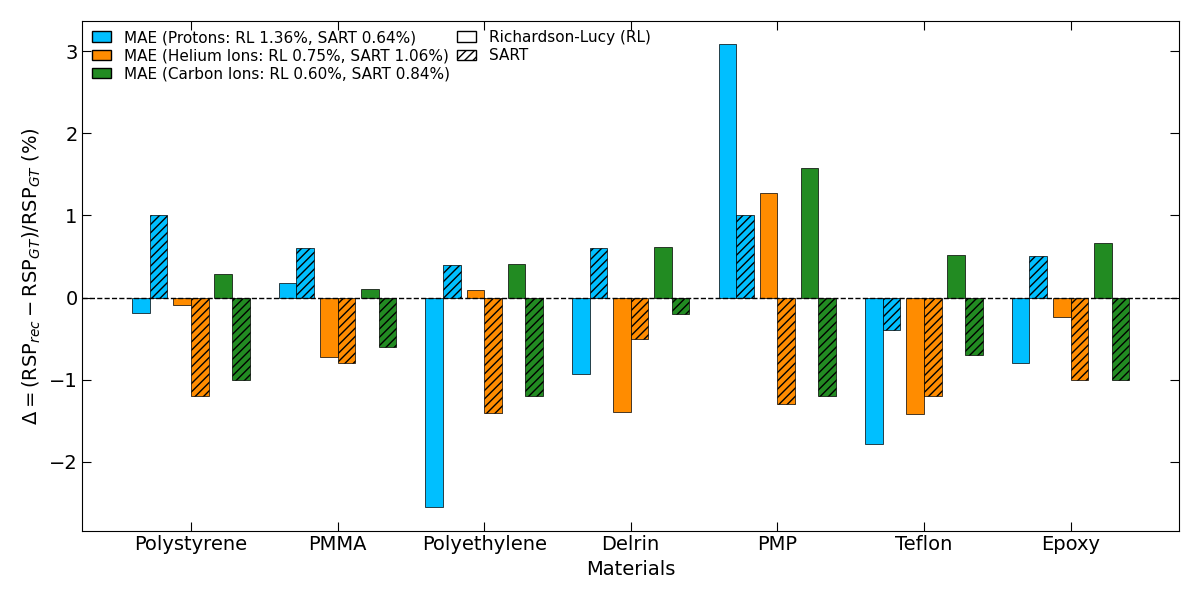}
    \caption{Relative differences between the ground truth and the reconstructed RSP values for all particles. In the legend the mean value of the modulus of the relative difference is indicated.
    }
    \label{fig:gt_vs_rec}
\end{figure}

Quantitative evaluation of the imaging performance was conducted by comparing the mean reconstructed RSP values within the regions of interest (ROIs) against the ground truth (GT) values derived from the GATE simulations. The detailed numerical comparison is summarized in Table \ref{tab:rsp_comparison}, while the relative differences ($\Delta = (RSP_{\text{rec}} - RSP_{\text{GT}}) / RSP_{\text{GT}}$) are illustrated in Figure \ref{fig:gt_vs_rec}.

\begin{table}[htbp]
\centering
\begin{tabular}{@{}lccccccc@{}}
\toprule
& & \multicolumn{2}{c}{\textbf{Proton}} & \multicolumn{2}{c}{\textbf{Helium}} & \multicolumn{2}{c}{\textbf{Carbon}} \\
\cmidrule(lr){3-4} \cmidrule(lr){5-6} \cmidrule(lr){7-8}
\textbf{Material} & \textbf{GT (Avg.)} & \textbf{RL} & \textbf{SART} & \textbf{RL} & \textbf{SART} & \textbf{RL} & \textbf{SART} \\
\midrule
    Polystyrene  & 1.047 & 1.045 & 1.054 & 1.047 & 1.031 & 1.049 & 1.033 \\
    PMMA         & 1.179 & 1.181 & 1.185 & 1.171 & 1.169 & 1.179 & 1.172 \\
    Polyethylene & 0.991 & 1.005 & 1.035 & 0.993 & 1.017 & 0.995 & 1.019 \\
    Delrin       & 1.364 & 1.352 & 1.375 & 1.347 & 1.360 & 1.371 & 1.363 \\
    PMP          & 0.863 & 0.890 & 0.901 & 0.876 & 0.880 & 0.875 & 0.881 \\
    Teflon       & 1.842 & 1.809 & 1.824 & 1.815 & 1.810 & 1.853 & 1.819 \\
    Epoxy        & 1.143 & 1.134 & 1.144 & 1.142 & 1.127 & 1.149 & 1.128 \\
\bottomrule
\end{tabular}
\caption{Comparison of the averaged Ground Truth (GT) and the Reconstructed RSP values using Richardson-Lucy (RL) and Simultaneous Algebraic Reconstruction Technique (SART) algorithms for Protons, Helium ions, and Carbon ions.}
\label{tab:rsp_comparison}
\end{table}

The analysis reveals material-dependent deviations characteristic of each particle type:

\textbf{Protons:} Proton imaging demonstrated varying degrees of accuracy across the material spectrum. When utilizing the Richardson-Lucy (RL) algorithm, reconstruction was exceptionally accurate for intermediate densities—matching the GT of PMMA almost perfectly ($+0.2\%$), but larger deviations were observed at the extremes. Specifically, RL reconstruction exhibited the highest discrepancy for the low-density PMP insert, with an overestimation of approximately $+3.1\%$. A similar trend was observed for Polyethylene, which was overestimated by $+1.5\%$. Conversely, the high-density Teflon insert was substantially underestimated by $-1.8\%$. When applying the SART algorithm, the overestimation for PMP and Polyethylene dropped to $+1.0\%$ and $+0.4\%$, respectively, while the Teflon underestimation was given $-0.4\%$. The mean value of the modulus of the relative difference was 1.2\% and 0.64\% in the case of RL and SART, respectively.

\textbf{Helium Ions:} Helium-based reconstruction showed a general tendency toward underestimation across most materials. With the RL solver, negative deviations were most pronounced for Teflon ($-1.4\%$) and Delrin ($-1.4\%$), as well as for the soft-tissue equivalent material PMMA ($-0.8\%$), though it reconstructed PMP with a much smaller overestimation error ($+1.3\%$) compared to protons. This systematic tendency toward underestimation was even more pronounced when applying the SART algorithm, which yielded negative deviations across all evaluated inserts (e.g., $-1.2\%$ for Teflon and $-1.3\%$ for PMP). The average of the modulus of the relative difference was given 0.75\% (RL) and 1.06\% (SART).

\textbf{Carbon Ions:} Carbon ion imaging displayed a highly consistent accuracy profile, demonstrating superior dosimetric stability independent of the chosen reconstruction algorithm. Under RL reconstruction, carbon ions showed a slight overall tendency toward overestimation, achieving the best accuracy for soft-tissue equivalent materials like PMMA ($+0.2\%$) and Polystyrene ($+0.3\%$). Conversely, the SART algorithm resulted in a tightly bounded, slight underestimation profile (e.g., $-0.6\%$ for PMMA and $-1.0\%$ for Polystyrene). A notable improvement was observed for the high-density Teflon insert regardless of the solver; unlike the lighter ions which underestimated this material, carbon ions achieved an error of only $+0.5\%$ with RL and $-0.7\%$ with SART. The MAE remained highly robust and consistent at 0.60\% for RL and 0.81\% for SART.

\begin{figure}[htbp]
    \centering
    \includegraphics[width=0.8\textwidth]{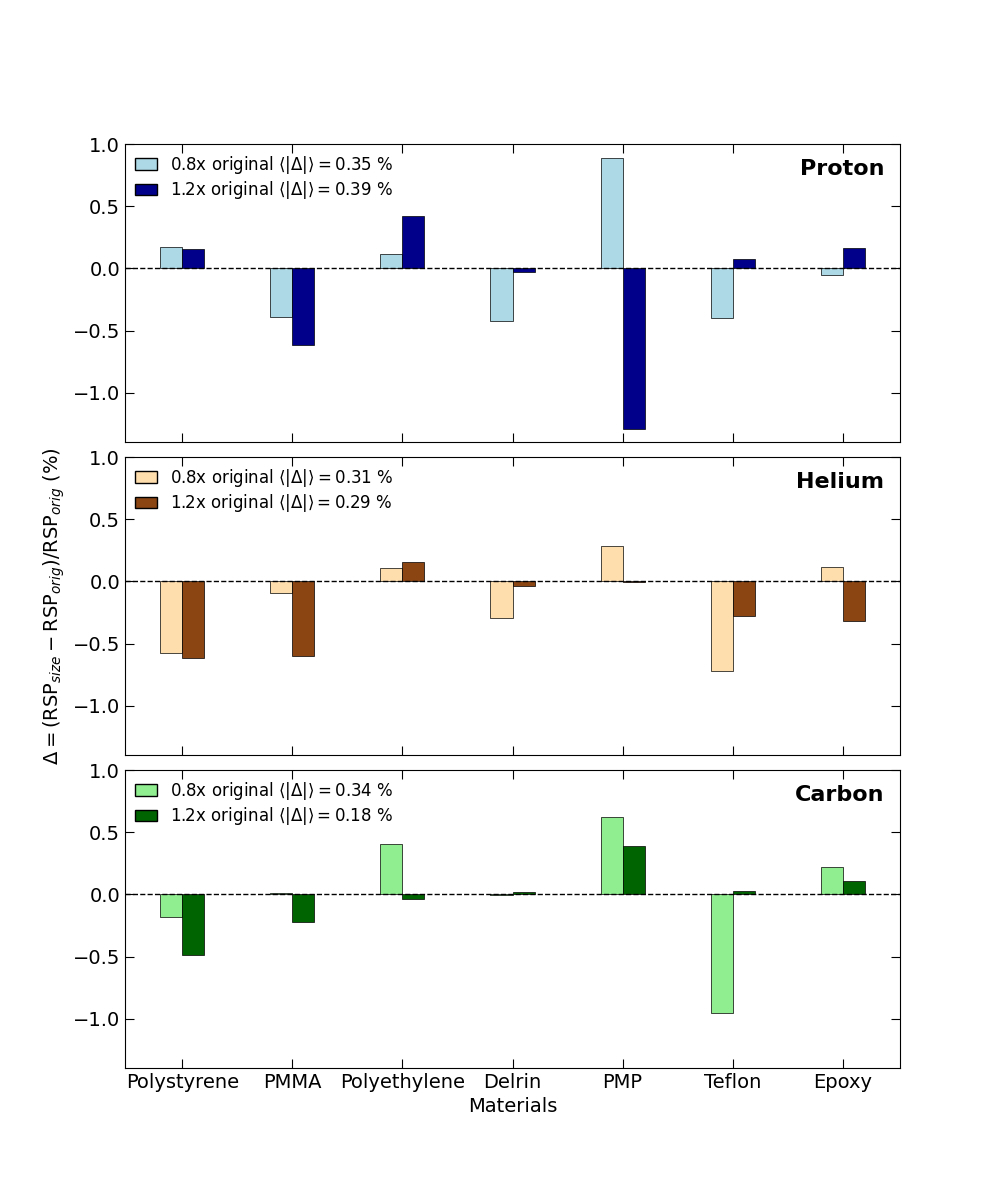}
    \caption{Differences of reconstructed RSP values between the original and the modified sized inserts.}
    \label{fig:sizes_diff}
\end{figure}

The evaluation of the RSP reconstruction accuracy was also extended to study the dependence on the size of the inserts in the phantom. The simulations were carried out using the previously described methods, using two modified CTP404 phantoms, where the inserts were 20\% smaller and 20\% larger. The reconstruction and the evaluation processes were also executed with the aforementioned methods.
The differences between the reconstructed RSP values of the modified phantoms and the original one can be seen on Figure \ref{fig:sizes_diff}. For the protons, except for the Polystyrene and the PMMA, an improvement is noticeable with the increasing insert size, however this effect is not present in the case of the helium ions and carbon ions. The difference between the RSP values of the original and the modified phantoms is within 1\% on average.

\subsection{Convergence and Dose Deposition}

\begin{figure}[htbp]
    \centering
    \includegraphics[width=0.8\textwidth]{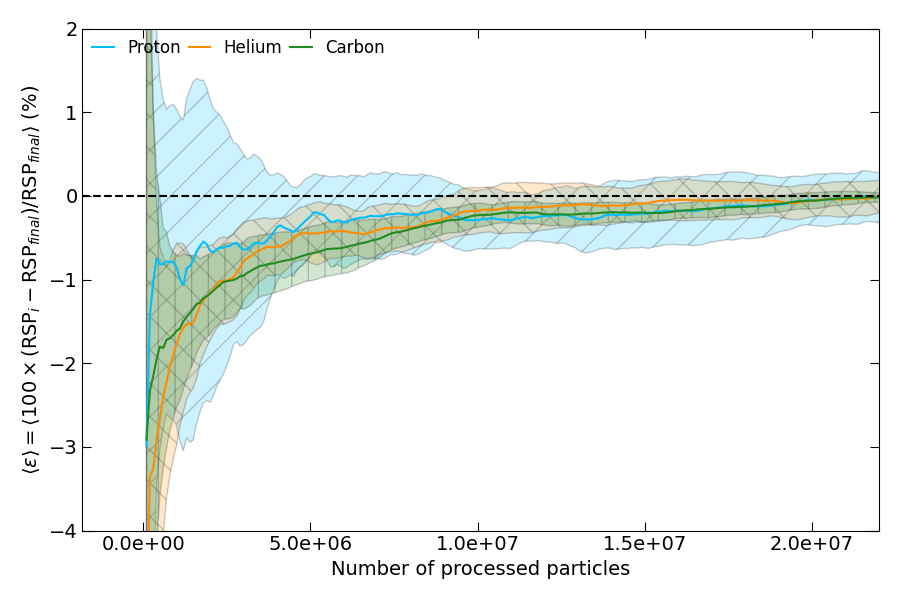}
    \caption{Relative difference between the ground truth reconstructed RSP values and the RSP values in the \textit{i}-th iteration of the reconstruction as a function of processed particles used for the reconstruction.}
    \label{fig:convergence}
\end{figure}

The analysis of the reconstruction convergence provides crucial insights into the minimum particle statistics required to achieve stable Relative Stopping Power (RSP) values. Figure \ref{fig:convergence} illustrates the relative difference between the RSP values at the $i$-th block and the ground truth reconstructed RSP (the final RSP values obtained after the previously described 1250th block and the Ruppert–Polyak averaging process) as a function of the number of processed particles. 

To systematically quantify the stability of the reconstruction, specific convergence thresholds were established. These thresholds are defined by the absolute value of the relative percentage difference ($|\epsilon|$) between the RSP at the $i$-th iteration and the ground truth reconstructed RSP:
\begin{equation}
    |\epsilon| = \left| \frac{\text{RSP}_i - \text{RSP}_{\text{GT}}}{\text{RSP}_{\text{GT}}} \right| \times 100\%.
\end{equation}
The particle counts required for the reconstruction curve to reach and strictly remain below these defined tolerance windows ($|\epsilon| \le 1.0\%$, $\le 0.5\%$, and $\le 0.25\%$), along with their corresponding doses for the CTP404 phantom, are detailed in Table \ref{tab:convergence_dose}.

\begin{table}[htbp]
  \centering
  \begin{tabular}{@{}clccc@{}}
    \toprule
    \textbf{Threshold} & \textbf{Metric} & \textbf{Proton} & \textbf{Helium} & \textbf{Carbon} \\
    \midrule
    \multirow{2}{*}{$\le 1.0\%$}  
    & Required Particles & 1,300,000 & 2,600,000 & 2,800,000 \\
    & Est. Dose (mGy)    & 0.003 & 0.022 & 0.130 \\
    \midrule
    \multirow{2}{*}{$\le 0.5\%$}  
    & Required Particles & 3,800,000 & 4,500,000 & 7,200,000 \\
    & Est. Dose (mGy)    & 0.009 & 0.038 & 0.333 \\
    \midrule
    \multirow{2}{*}{$\le 0.25\%$} 
    & Required Particles & 13,800,000 & 9,200,000 & 9,800,000 \\
    & Est. Dose (mGy)    & 0.032 & 0.078 & 0.454 \\
    \bottomrule
  \end{tabular}
  \caption{Number of processed particles required to reach and maintain absolute difference convergence thresholds relative to the final reconstructed RSP, alongside the corresponding estimated imaging dose deposited in the CTP404 phantom.}
\label{tab:convergence_dose}
\end{table}

Because the convergence analysis relies on tracking the structural resolution and the evolution of RSP reconstruction accuracy of specific density inserts, it was performed directly on the CTP404 phantom. Since energy deposition scales linearly with the number of simulated primary particles, the convergence thresholds detailed in Table \ref{tab:convergence_dose} correspond to specific fractions of the total 25 million particle dataset. For protons, reaching the $1.0\%$, $0.5\%$ and $0.25\%$ convergence thresholds required 5.2\%, 15.2\%, and 55.2\% of the total simulated dose, respectively. For helium ions, these thresholds corresponded to 10.4\%, 18.0\%, and 36.8\%, while carbon ions required 11.2\%, 28.8\%, and 39.2\% of their respective total doses.

The convergence profiles reveal distinct physical behaviors for the different particle species. While protons rapidly reach loose convergence thresholds (requiring only $1.3 \times 10^6$ particles to fall within $1.0\%$ of the final RSP), their convergence curve exhibits significant statistical fluctuation due to higher multiple Coulomb scattering. Consequently, protons require the highest number of particles ($13.8 \times 10^6$) to fully stabilize within the strict $\le 0.25\%$ threshold. In contrast, heavier ions display smoother, more monotonic convergence behaviors due to their stiffer particle trajectories. To achieve the high-fidelity $\le 0.25\%$ threshold, helium and carbon ions require only $9.2 \times 10^6$ and $9.8 \times 10^6$ particles, respectively. This corresponds to a statistical requirement of approximately $67\%$ and $71\%$ of the particles needed for protons to achieve the same strict stability.

Despite requiring fewer particles for strict convergence, Table \ref{tab:convergence_dose} demonstrates that carbon and helium ions still deliver a higher absolute dose than protons at the $\le 0.25\%$ threshold. This escalation in deposited dose is driven by several underlying physical factors. First, according to the Bethe-Bloch equation, electronic stopping power scales with the square of the particle's effective charge ($Z^2$). Consequently, heavier ions naturally exhibit a substantially higher linear energy transfer (LET) compared to protons (\cite{schardt2010heavy}). Second, to achieve an equivalent penetration depth in the phantom, the carbon ions required a higher initial kinetic energy ($430$ MeV/u) compared to the protons and helium ions ($230$ MeV/u), meaning significantly more total energy was introduced into the system. Finally, unlike protons, heavier ions—particularly carbon—undergo inelastic nuclear collisions as they traverse the medium. This fragmentation process creates a spectrum of lighter, secondary particles (such as protons and alpha particles) that continue to deposit energy even beyond the primary Bragg peak, contributing to a higher integral dose (\cite{gunzert2008secondary}).

To provide a clinically realistic estimation of the radiation, we applied these identical convergence fractions to the 3-dimensional Shepp-Logan-like head phantom. The fully 3-dimensional imaging resulted in approximately $456\times10^6$ primary particles, meaning a mean dose of 0.97 mGy, 3.57 mGy, and 19.11 mGy for protons, helium ions and carbon ions, respectively. Extrapolating the convergence fractions to this clinical geometry, achieving a stable 1\% RSP convergence threshold would result in an estimated human head dose of only 0.050 mGy for protons, 0.372 mGy for helium, and 2.140 mGy for carbon ions. Reaching the strictest 0.25\% threshold would necessitate corresponding head doses of 0.534 mGy, 1.316 mGy, and 7.490 mGy for protons, helium, and carbon ions, respectively. The results are summarized in Table \ref{tab:sl_dose_extrapolation}

\begin{table}[htbp]
  \centering
  \begin{tabular}{@{}clccc@{}}
    \toprule
    \textbf{Threshold} & \textbf{Metric} & \textbf{Proton} & \textbf{Helium} & \textbf{Carbon} \\
    \midrule
    \multirow{1}{*}{Full Scan} 
    & Total Dose (mGy) & 0.967 & 3.575 & 19.106 \\
    \midrule
    \multirow{2}{*}{$\le 1.0\%$}  
    & Dose Fraction    & 5.2\% & 10.4\% & 11.2\% \\
    & Est. Dose (mGy)  & 0.050 & 0.372 & 2.140 \\
    \midrule
    \multirow{2}{*}{$\le 0.5\%$}  
    & Dose Fraction    & 15.2\% & 18.0\% & 28.8\% \\
    & Est. Dose (mGy)  & 0.147 & 0.643 & 5.503 \\
    \midrule
    \multirow{2}{*}{$\le 0.25\%$} 
    & Dose Fraction    & 55.2\% & 36.8\% & 39.2\% \\
    & Est. Dose (mGy)  & 0.534 & 1.316 & 7.490 \\
    \bottomrule
  \end{tabular}
  \caption{Estimated imaging dose deposited in the 3D Shepp-Logan-like phantom. The fractional doses correspond to the percentage of particles required to reach the specific RSP convergence thresholds established in the CTP404 phantom.}
\label{tab:sl_dose_extrapolation}
\end{table}

\section{Discussion and Comparison} 
\label{sec:discussion}
\subsection{Dosimetric Performance}
One of the most significant findings of this study is the substantial dose reduction offered by particle imaging modalities compared to conventional X-ray computed tomography. As shown in Table \ref{tab:sl_dose_extrapolation}, the estimated absorbed dose for a 3-dimensional head-like phantom scan, taking strict reconstruction convergence requirements (0.25\%) into account, can be reduced to as low as $0.534$ mGy, $1.316$ mGy, and $7.490$ mGy, for protons, helium ions, and carbon ions, respectively.

These values are markedly lower than standard clinical protocols. A typical diagnostic head CT scan delivers an organ absorbed dose to the brain approximately 40 mGy (\cite{huda2007patient}). Even the carbon-ion scan yielded a dose over five times lower than a standard X-ray CT at the strictest convergence threshold. The proton and helium scans offer an even more dramatic reduction, potentially lowering the dose by a factor of approximately $30$ to over $75$ compared to photon-based imaging. It is worth noting that while carbon ions impart a higher physical dose than protons due to their higher Linear Energy Transfer (LET), this is accompanied by reduced multiple Coulomb scattering, which theoretically supports higher spatial resolution (\cite{gehrke2018theoretical}).

The absolute results can vary based on the imaging protocol being utilized; however they provide a reliable estimation demonstrating that this characteristic could be particularly advantageous for pediatric patients or patients requiring frequent re-imaging for adaptive therapy, where cumulative radiation exposure is a primary concern.

\subsection{Comparative Imaging Accuracy}
The reconstruction accuracy achieved in this work (typically within $1-2\%$) is consistent with the current state-of-the-art in particle imaging research and compares favorably with X-ray CT benchmarks.

In clinical practice, stoichiometric calibration of single-energy CT (SECT) typically yields RSP uncertainties in the range of $1.6\%$ to $3.5\%$ (\cite{hansen2015simulation, yang2012comprehensive}). Dual-energy CT (DECT) has improved this significantly, achieving accuracies of approximately $0.5\%$ to $1\%$ (\cite{dedes2019experimental}).

Our estimated dose depositions are consistent with the results of \cite{https://doi.org/10.1002/mp.15283}, who reported estimated doses of approximately 4.7~mGy and 6.6~mGy for pCT and HeCT, respectively. These values represent a reduction by a factor of 9 compared to SECT and a factor of 13 compared to DECT, maintaining the same order of magnitude as the dosimetric savings observed in our study.

Our results indicate that carbon-ion and helium-ion imaging could achieve RSP accuracy competitive with DECT and superior to SECT, without the need for empirical photon-to-hadron conversion models. While exceptional accuracy was observed for specific tissue-equivalent materials (e.g., approximately 0.2\% error for PMMA with carbon ions under RL reconstruction), the imaging fidelity remained highly robust across the entire investigated spectrum and across both mathematical solvers. For carbon-ion imaging, the relative RSP error was maintained well below 1\% for the vast majority of the inserts, with maximum deviations strictly bounded at 1.6\% (RL) and 1.2\% (SART) for the lowest density materials. Similarly, helium-ion imaging demonstrated consistent precision, keeping relative errors strictly below 1.5\% regardless of the applied reconstruction algorithm. This generalized stability across varying densities, compositions and mathematical approaches strongly underscores the reliable clinical imaging potential of these heavy-ion modalities.

The proton imaging results (errors up to $3.1\%$ for PMP under RL reconstruction) show slightly higher deviations than the $1-2\%$ accuracy reported in some idealized simulation studies (\cite{schulte2005density}). While the application of the additive SART algorithm effectively suppressed the $+3.1\%$ deviation to $+1.0\%$ (suggesting that the high deviation was primarily an algorithmic artifact rooting from the multiplicative update rule of the RL solver), the baseline susceptibility of protons to scattering artifacts remains evident. This is likely attributable to the simplified Most Likely Path (MLP) formalism used here, which may not fully account for the complex scattering in heterogeneous media compared to more advanced probabilistic solvers. However, our results for helium and carbon ions align well with recent experimental findings. For instance, Gehrke et al. reported helium CT relative range errors of roughly $0.3\%$ to $0.7\%$ (\cite{gehrke2018theoretical}), which supports our observation that heavier ions can provide robust reconstruction fidelity.

The evaluation the high-density Teflon insert highlights the non-trivial relationship between particle species and energy loss models in high-density materials. While lighter ions exhibited substantial underestimations under multiplicative RL updates ($-1.8\%$ for protons, $-1.4\%$ for helium ions), carbon ions maintained a highly stable estimate ($+0.5\%$ with RL, $-0.7\%$ with SART). Furthermore, applying the additive SART solver significantly reduced the proton underestimation in Teflon from $-1.8\%$ to $-0.4\%$, demonstrating that a large portion of this discrepancy is an algorithmic artifact caused by the RL solver struggling with steep, high-density gradients. However, a baseline physical uncertainty remains. As established by other Monte Carlo studies (\cite{doolan2016inter}), intrinsic simulation parameters (such as nuclear interaction models and ionization potentials within toolkits like Geant4) are recognized as critical constraints that fundamentally affect the absolute accuracy of heavy-ion transport and RSP reconstruction. 

In conclusion, while carbon ions deliver a higher imaging dose than protons or helium, they remain well below the dose burden of conventional X-ray CT while offering better RSP accuracy and faster statistical convergence for standard tissue-equivalent materials. Helium ions appear to offer a balanced compromise, providing better accuracy than protons (due to reduced scattering) with a dose footprint significantly lower than carbon ions.

While this study clearly demonstrates the dosimetric and RSP reconstruction advantages of heavy-ion CT, we acknowledge certain methodological limitations. First, our simulations presupposed an ideal detector system to strictly isolate the fundamental physical interactions of the ions. The implementation of realistic detector uncertainties, electronic noise, and physical hardware limitations will be necessary to fully translate these findings to a clinical setup. Furthermore, while improved spatial resolution is a primary motivation for utilizing heavier ions (due to their reduced multiple Coulomb scattering) a rigorous quantitative assessment of image resolution, such as the evaluation of the Modulation Transfer Function (MTF), falls outside the scope of this specific investigation. The current study was strictly designed to evaluate the baseline dosimetric footprint and statistical convergence of RSP recovery. A comprehensive quantitative evaluation of spatial resolution metrics, alongside realistic detector modeling, will form the primary focus of our upcoming follow-up studies.

\section{Summary}
\label{sec:summary} 

In this work, we presented a comprehensive comparative study of proton, helium-ion, and carbon-ion computed tomography (CT) for charged-particle therapy treatment planning. Using high-fidelity Monte Carlo simulations (GATE/Geant4), we evaluated the relative stopping power (RSP) reconstruction accuracy, statistical convergence behavior, and the associated imaging dose for a standard CTP404 phantom.

Our results demonstrate that direct particle imaging offers a viable and potentially superior alternative to conventional X-ray CT for calculating stopping power maps. The key findings are summarized as follows:

\textbf{Dosimetric Advantage:} All three particle imaging modalities demonstrated a substantial reduction in radiation burden compared to standard clinical protocols ($\sim40$ mGy). When optimized for strict statistical convergence, carbon-ion imaging requires an estimated dose of $\sim7.5$ mGy for a human head equivalent. Proton and helium-ion imaging offer further reductions, minimizing the imaging dose to practically negligible levels ($\sim0.5$ mGy and $\sim1.3$ mGy, respectively).

\textbf{Reconstruction Accuracy:} Carbon ions yielded reassuring reconstruction fidelity for soft-tissue equivalent materials (e.g., PMMA, Polystyrene) and high-density materials (e.g., Teflon), achieving absolute errors generally well below 1.0\% across both the RL and SART reconstruction algorithms. This promising performance is attributed to the reduced multiple Coulomb scattering of heavier ions, which preserves spatial information more effectively than protons. Helium ions served as an effective compromise, offering better accuracy than protons in low-density regions (such as PMP) while maintaining a significantly lower dose profile than carbon ions. Protons, while delivering the absolute lowest dose, exhibited the highest range of RSP deviations, particularly in materials with extreme densities. Although utilizing the additive SART algorithm successfully mitigated the most extreme proton RSP deviations (reducing the maximum observed error for PMP form $+3.1\%$ down to $1.0\%$), protons consistently displayed lower physical robustness compared to heavier ions. The RSP reconstruction accuracy did not show a strong dependence on the insert size, as the difference between the RSP values of the different sized phantoms did not exceed $1\%$ on average.

\textbf{Statistical Convergence:} Heavy ions demonstrated promising spatial stability during the iterative reconstruction process. Carbon and helium ions achieved strict convergence thresholds ($\le 0.25\%$ absolute difference) requiring approximately $29\%$ and $33\%$ fewer particles than protons, respectively, further validating their efficiency for high-fidelity imaging.

\textbf{Clinical Implications:} The trade-off between image noise, spatial resolution, and patient dose suggests that the optimal particle species may depend on the specific clinical scenario. For pediatric cases or frequent adaptive replanning, proton or helium imaging provides exceptional dose sparing. Conversely, for detailed treatment planning in heterogeneous regions where range accuracy is paramount, carbon-ion imaging offers the highest precision and fastest statistical stability.

Furthermore, this study demonstrates that an algorithm not specifically designed for ion applications can nonetheless perform effective and clinically relevant dose planning using ion beams, provided that accurate ion detection is achieved.

It is worth noting that our study and investigations were executed under idealized circumstances. The results yield a good estimation for the dose deposition and RSP reconstruction accuracy, however future work will focus on validating these simulation results with experimental beam data and investigating the impact of these RSP improvements on the final dose distribution in patient-specific treatment plans.

\section*{Acknowledgement}

The authors would like to thank the support of the EKÖP-25 University Research Scholarship Program of the Ministry for Culture and Innovation from the source of the National Research, Development and Innovation Fund. Furthermore, the authors acknowledge ELTE Eötvös Loránd University, Budapest, Hungary, as a supporting institution. This work was supported by the Hungarian National Research, Development and Innovation Office (NKFIH) grants under the contract numbers NKKP ADVANCED 25-153456, 2025-1.1.5-NEMZ\_KI-2025-00005, 2025-1.1.5-NEMZ\_KI-2025-00013, 2024-1.2.5-TÉT-2024-00022 and the FuSe COST Action CA-24101. Computational resources were provided by the Wigner Scientific Computing Laboratory (WSCLAB) and the HUN-REN Wigner Cloud (see \cite{heder2022past}; \url{https://science-cloud.hu/}).

\section*{Members of the Bergen pCT Collaboration}

\footnotesize
\begin{flushleft}
Johan Alme\textsuperscript{1}, Gergely Gábor Barnaföldi\textsuperscript{2}, Gábor Bíró\textsuperscript{2}, Tea Bodova\textsuperscript{1,3}, Illia Borschov\textsuperscript{5}, Vyacheslav Borshchov\textsuperscript{5}, Anthony van den Brink\textsuperscript{1}, Lattawat Charoonratana\textsuperscript{1}, Bence Dudás\textsuperscript{4}, Gregory Feofilov\textsuperscript{7}, Nicolas R. Gauger\textsuperscript{8}, Håvard Helstrup\textsuperscript{9}, Sergey Igolkin\textsuperscript{7}, Zsófia Jólesz\textsuperscript{2,4}, Ralf Keidel\textsuperscript{8}, Thitayawat Khumwong\textsuperscript{6}, Chinorat Kobdaj\textsuperscript{6}, Tobias Kortus\textsuperscript{8}, Natthawut Laojamnongwong\textsuperscript{6}, Odd Harald Odland\textsuperscript{1,10}, Gábor Papp\textsuperscript{4}, Thomas Peitzmann\textsuperscript{11}, Helge Egil Seime Pettersen\textsuperscript{10}, Maksym Protsenko\textsuperscript{5}, Attiq Ur Rehman\textsuperscript{1}, Matthias Richter\textsuperscript{1}, Narongrit Ritjoho\textsuperscript{6}, Dieter Röhrich\textsuperscript{1}, Hira Sharif\textsuperscript{12}, Alexander Schilling\textsuperscript{8}, Tawanchat Simantathammakul\textsuperscript{6}, Ihor Tymchuk\textsuperscript{5}, Kjetil Ullaland\textsuperscript{1}, Shiming Yang\textsuperscript{1}, Jie Yi\textsuperscript{13}

\vspace{0.4cm}

1) Department of Physics and Technology, University of Bergen, 5007 Bergen, Norway; 
2) HUN-REN Wigner Research Centre for Physics, 29--33 Konkoly--Thege Mikl\'os \'ut, H-1121 Budapest, Hungary;
3) European Organization for Nuclear Research (CERN), Geneva, Switzerland
4) Institute for Physics, Eötvös Loránd University, 1/A Pázmány P. Sétány, H-1117 Budapest, Hungary; 
5) Research and Production Enterprise "LTU" (RPELTU), Kharkiv, Ukraine; 
6) Institute of Science, Suranaree University of Technology, Nakhon Ratchasima, Thailand; 
7) St. Petersburg University, St. Petersburg, Russia; 
8) Chair for Scientific Computing, RPTU University Kaiserslautern-Landau, 67663 Kaiserslautern, Germany; 
9) Department of Computer Science, Electrical Engineering and Mathematical Sciences, Western Norway University of Applied Sciences, 5020 Bergen, Norway; 
10) Cancer Clinic, Haukeland University Hospital, 5021 Bergen, Norway; 
11) Institute for Gravitational and Subatomic Physics (GRASP), Utrecht University, Utrecht, Netherlands; 
12) University of Jyväskylä, Jyväskylä, Finland; 
13) Central China Normal University, Wuhan, China
\end{flushleft}
\normalsize

\printbibliography

@article{payandeh2025comparative,
  title={A comparative analysis of the proton, helium-ion, and carbon-ion therapy for brain tumors},
  author={Payandeh, Sina and Keshavarz, Sajad and Malidarreh, Roya Boudaghi and Yasamani, Fatemeh Hasanzadeh},
  journal={Physica Scripta},
  volume={100},
  number={8},
  pages={085308},
  year={2025},
  publisher={IOP Publishing}
}

@article{wickert2022radiotherapy,
  title={Radiotherapy with Helium ions has the potential to Improve both Endocrine and Neurocognitive Outcome in Pediatric patients with Ependymoma},
  author={Wickert, Ricarda and Tessonnier, Thomas and Deng, Maximilian and Adeberg, Sebastian and Seidensaal, Katharina and Hoeltgen, Line and Debus, J{\"u}rgen and Herfarth, Klaus and Harrabi, Semi B},
  journal={Cancers},
  volume={14},
  number={23},
  pages={5865},
  year={2022},
  publisher={MDPI}
}

@article{huang2022comparison,
  title={Comparison of the efficacy and toxicity of postoperative proton versus carbon ion radiotherapy for head and neck cancers},
  author={Huang, Qingting and Hu, Jiyi and Hu, Weixu and Gao, Jing and Yang, Jing and Qiu, Xianxin and Lu, Jiade Jay and Kong, Lin},
  journal={Annals of Translational Medicine},
  volume={10},
  number={22},
  pages={1197},
  year={2022}
}

@article{kaiser2019proton,
  title={Proton therapy delivery and its clinical application in select solid tumor malignancies},
  author={Kaiser, Adeel and Eley, John G and Onyeuku, Nasarachi E and Rice, Stephanie R and Wright, Carleen C and McGovern, Nathan E and Sank, Megan and Zhu, Mingyao and Vujaskovic, Zeljko and Simone 2nd, Charles B and others},
  journal={Journal of Visualized Experiments (JoVE)},
  number={144},
  pages={e58372},
  year={2019}
}

@article{ordonez2019fast,
  title={Fast in situ image reconstruction for proton radiography},
  author={Ordo{\~n}ez, Caesar E and Karonis, Nicholas T and Duffin, Kirk L and Winans, John R and DeJongh, Ethan A and DeJongh, Don F and Coutrakon, George and Myers, Nicole F and Pankuch, Mark and Welsh, James S},
  journal={Journal of radiation oncology},
  volume={8},
  number={2},
  pages={185--198},
  year={2019},
  publisher={Springer}
}

@article{hansen2015simulation,
  title={A simulation study on proton computed tomography (CT) stopping power accuracy using dual energy CT scans as benchmark},
  author={Hansen, David C and Seco, Joao and S{\o}rensen, Thomas Sangild and Petersen, J{\o}rgen Breede Baltzer and Wildberger, Joachim E and Verhaegen, Frank and Landry, Guillaume},
  journal={Acta oncologica},
  volume={54},
  number={9},
  pages={1638--1642},
  year={2015},
  publisher={Taylor \& Francis}
}

@misc{CTP404,
  author = {{The Phantom Laboratory}},
  title = {Catphan® 600 phantom (containing CTP404 and CTP528 as a section)},
  year = {2022},
  note = {Last accessed 5 October 2022},
  url = {https://www.phantomlab.com/catphan-600}
}

@ARTICLE{Allison2006,
  author={Allison, J. and Amako, K. and Apostolakis, J. and Araujo, H. and Arce Dubois, P. and Asai, M. and Barrand, G. and Capra, R. and Chauvie, S. and Chytracek, R. and Cirrone, G. A. P. and Cooperman, G. and Cosmo, G. and Cuttone, G. and Daquino, G. G. and Donszelmann, M. and Dressel, M. and Folger, G. and Foppiano, F. and Generowicz, J. and Grichine, V. and Guatelli, S. and Gumplinger, P. and Heikkinen, A. and Hrivnacova, I. and Howard, A. and Incerti, S. and Ivanchenko, V. and Johnson, T. and Jones, F. and Koi, T. and Kokoulin, R. and Kossov, M. and Kurashige, H. and Lara, V. and Larsson, S. and Lei, F. and Link, O. and Longo, F. and Maire, M. and Mantero, A. and Mascialino, B. and McLaren, I. and Mendez Lorenzo, P. and Minamimoto, K. and Murakami, K. and Nieminen, P. and Pandola, L. and Parlati, S. and Peralta, L. and Perl, J. and Pfeiffer, A. and Pia, M.G. and Ribon, A. and Rodrigues, P. and Russo, G. and Sadilov, S. and Santin, G. and Sasaki, T. and Smith, D. and Starkov, N. and Tanaka, S. and Tcherniaev, E. and Tome, B. and Trindade, A. and Truscott, P. and Urban, L. and Verderi, M. and Walkden, A. and Wellisch, J. P. and Williams, D. C. and Wright, D. and Yoshida, H.},
  journal={IEEE Transactions on Nuclear Science}, 
  title={Geant4 developments and applications}, 
  year={2006},
  volume={53},
  number={1},
  pages={270-278},
  doi={10.1109/TNS.2006.869826}}

@article{Agostinelli2003,
title = {Geant4—a simulation toolkit},
journal = {Nuclear Instruments and Methods in Physics Research Section A: Accelerators, Spectrometers, Detectors and Associated Equipment},
volume = {506},
number = {3},
pages = {250-303},
year = {2003},
issn = {0168-9002},
doi = {https://doi.org/10.1016/S0168-9002(03)01368-8},
author = {S. Agostinelli and J. Allison and K. Amako and J. Apostolakis and H. Araujo and P. Arce and M. Asai and D. Axen and S. Banerjee and G. Barrand and F. Behner and L. Bellagamba and J. Boudreau and L. Broglia and A. Brunengo and H. Burkhardt and S. Chauvie and J. Chuma and R. Chytracek and G. Cooperman and G. Cosmo and P. Degtyarenko and A. Dell'Acqua and G. Depaola and D. Dietrich and R. Enami and A. Feliciello and C. Ferguson and H. Fesefeldt and G. Folger and F. Foppiano and A. Forti and S. Garelli and S. Giani and R. Giannitrapani and D. Gibin and J. J. {Gómez Cadenas} and I. González and G. {Gracia Abril} and G. Greeniaus and W. Greiner and V. Grichine and A. Grossheim and S. Guatelli and P. Gumplinger and R. Hamatsu and K. Hashimoto and H. Hasui and A. Heikkinen and A. Howard and V. Ivanchenko and A. Johnson and F.W. Jones and J. Kallenbach and N. Kanaya and M. Kawabata and Y. Kawabata and M. Kawaguti and S. Kelner and P. Kent and A. Kimura and T. Kodama and R. Kokoulin and M. Kossov and H. Kurashige and E. Lamanna and T. Lampén and V. Lara and V. Lefebure and F. Lei and M. Liendl and W. Lockman and F. Longo and S. Magni and M. Maire and E. Medernach and K. Minamimoto and P. {Mora de Freitas} and Y. Morita and K. Murakami and M. Nagamatu and R. Nartallo and P. Nieminen and T. Nishimura and K. Ohtsubo and M. Okamura and S. O'Neale and Y. Oohata and K. Paech and J. Perl and A. Pfeiffer and M.G. Pia and F. Ranjard and A. Rybin and S. Sadilov and E. {Di Salvo} and G. Santin and T. Sasaki and N. Savvas and Y. Sawada and S. Scherer and S. Sei and V. Sirotenko and D. Smith and N. Starkov and H. Stoecker and J. Sulkimo and M. Takahata and S. Tanaka and E. Tcherniaev and E. {Safai Tehrani} and M. Tropeano and P. Truscott and H. Uno and L. Urban and P. Urban and M. Verderi and A. Walkden and W. Wander and H. Weber and J. P. Wellisch and T. Wenaus and D. C. Williams and D. Wright and T. Yamada and H. Yoshida and D. Zschiesche},
}

@article{Jan2004,
doi	= {10.1088/0031-9155/49/19/007},
title	= {GATE: a simulation toolkit for PET and SPECT},
author	= {Jan, S. and Santin, G. and Strul, D. and Staelens, S. and Assié, K. and Autret, D. and Avner, S. and Barbier, R. and Bardiès, M. and Bloomfield, P. M.},
journal	= {Physics in Medicine and Biology    2004-sep 10 vol. 49 iss. 19},

year	= {2004},
month	= {sep},
day	= {10},
volume	= {49},
issue	= {19},
page	= {4543--4561}
}

@article{Jan2011,
doi	= {10.1088/0031-9155/56/4/001},
title	= {GATE V6: a major enhancement of the GATE simulation platform enabling modelling of CT and radiotherapy},
author	= {Jan, S. and Benoit, D. and Becheva, E. and Carlier, T. and Cassol, F. and Descourt, P. and Frisson, T. and Grevillot, L. and Guigues, L. and Maigne, L. and Morel, C. and Perrot, Y. and Rehfeld, N. and Sarrut, D. and Schaart, D. R. and Stute, S. and Pietrzyk, U. and Visvikis, D. and Zahra, N. and Buvat, I.},
journal	= {Physics in Medicine and Biology    2011-jan 20 vol. 56 iss. 4},

year	= {2011},
month	= {jan},
day	= {20},
volume	= {56},
issue	= {4},
page	= {881--901}
}

@article{gehrke2018theoretical,
  title={Theoretical and experimental comparison of proton and helium-beam radiography using silicon pixel detectors},
  author={Gehrke, T and Amato, C and Berke, S and Marti{\v{s}}{\'\i}kov{\'a}, M},
  journal={Physics in Medicine \& Biology},
  volume={63},
  number={3},
  pages={035037},
  year={2018},
  publisher={IOP Publishing}
}

@article{yang2012comprehensive,
  title={Comprehensive analysis of proton range uncertainties related to patient stopping-power-ratio estimation using the stoichiometric calibration},
  author={Yang, Ming and Zhu, X Ronald and Park, Peter C and Titt, Uwe and Mohan, Radhe and Virshup, Gary and Clayton, James E and Dong, Lei},
  journal={Physics in Medicine \& Biology},
  volume={57},
  number={13},
  pages={4095},
  year={2012},
  publisher={IOP Publishing}
}

@article{dedes2019experimental,
  title={Experimental comparison of proton CT and dual energy x-ray CT for relative stopping power estimation in proton therapy},
  author={Dedes, George and Dickmann, Jannis and Niepel, Katharina and Wesp, Philipp and Johnson, Robert P and Pankuch, Mark and Bashkirov, Vladimir and Rit, Simon and Volz, Lennart and Schulte, Reinhard W and others},
  journal={Physics in Medicine \& Biology},
  volume={64},
  number={16},
  pages={165002},
  year={2019},
  publisher={IOP Publishing}
}

@article{schulte2005density,
  title={Density resolution of proton computed tomography},
  author={Schulte, Reinhard W and Bashkirov, Vladimir and Loss Klock, M{\'a}rgio C and Li, Tianfang and Wroe, Andrew J and Evseev, Ivan and Williams, David C and Satogata, Todd},
  journal={Medical physics},
  volume={32},
  number={4},
  pages={1035--1046},
  year={2005},
  publisher={Wiley Online Library}
}

@article{doolan2016inter,
  title={Inter-comparison of relative stopping power estimation models for proton therapy},
  author={Doolan, PJ and Collins-Fekete, Charles-Antoine and Dias, Marta F and Ruggieri, Thomas A and D’Souza, Derek and Seco, Joao},
  journal={Physics in Medicine \& Biology},
  volume={61},
  number={22},
  pages={8085},
  year={2016},
  publisher={IOP Publishing}
}

@article{borges2023pros,
  title={Pros and cons of dual-energy CT systems:“one does not fit all”},
  author={Borges, Ana P and Antunes, C{\'e}lia and Curvo-Semedo, Lu{\'\i}s},
  journal={Tomography},
  volume={9},
  number={1},
  pages={195--216},
  year={2023},
  publisher={MDPI}
}

@article{giordanengo2015cnao,
  title={The CNAO dose delivery system for modulated scanning ion beam radiotherapy},
  author={Giordanengo, Simona and Garella, Maria Adelaide and Marchetto, Flavio and Bourhaleb, Faiza and Ciocca, Mario and Mirandola, Alfredo and Monaco, Vincenzo and Hosseini, Mohammad Amin and Peroni, Cristiana and Sacchi, Roberto and others},
  journal={Medical physics},
  volume={42},
  number={1},
  pages={263--275},
  year={2015},
  publisher={Wiley Online Library}
}

@article{piersimoni2018helium,
  title={Helium CT: Monte Carlo simulation results for an ideal source and detector with comparison to proton CT},
  author={Piersimoni, Pierluigi and Faddegon, Bruce A and M{\'e}ndez, Jos{\'e} Ramos and Schulte, Reinhard W and Volz, Lennart and Seco, Joao},
  journal={Medical physics},
  volume={45},
  number={7},
  pages={3264--3274},
  year={2018},
  publisher={Wiley Online Library}
}

@misc{bíró2026protoncomputedtomographyimage,
      title={Proton Computed Tomography Image Reconstruction Based on the Richardson-Lucy Algorithm}, 
      author={Gábor Bíró and Ákos Sudár and Zsófia Jólesz and Gábor Papp and Gergely Gábor Barnaföldi},
      year={2026},
      eprint={2212.00126},
      archivePrefix={arXiv},
      primaryClass={physics.med-ph},
      url={https://arxiv.org/abs/2212.00126}, 
}

@article{schardt2010heavy,
  title={Heavy-ion tumor therapy: Physical and radiobiological benefits},
  author={Schardt, Dieter and Els{\"a}sser, Thilo and Schulz-Ertner, Daniela},
  journal={Reviews of modern physics},
  volume={82},
  number={1},
  pages={383--425},
  year={2010},
  publisher={APS}
}

@article{gunzert2008secondary,
  title={Secondary beam fragments produced by 200 MeV u- 1 12C ions in water and their dose contributions in carbon ion radiotherapy},
  author={Gunzert-Marx, K and Iwase, H and Schardt, D and Simon, et RS},
  journal={New journal of physics},
  volume={10},
  number={7},
  pages={075003},
  year={2008}
}

@article{rit2013filtered,
  title={Filtered backprojection proton CT reconstruction along most likely paths},
  author={Rit, Simon and Dedes, George and Freud, Nicolas and Sarrut, David and L{\'e}tang, Jean Michel},
  journal={Medical physics},
  volume={40},
  number={3},
  pages={031103},
  year={2013},
  publisher={Wiley Online Library}
}

@article{penfold2010block,
  title={Block-iterative and string-averaging projection algorithms in proton computed tomography image reconstruction},
  author={Penfold, Scott N and Schulte, Reinhard W and Censor, Yair and Bashkirov, Vladimir and McAllister, Scott and Schubert, Keith E and Rosenfeld, Anatoly B},
  journal={Biomedical Mathematics: Promising Directions in Imaging, Therapy Planning and Inverse Problems},
  pages={347--367},
  year={2010},
  publisher={Medical Physics Publishing Madison, WI}
}

@article{DesignPixelRangeTelescope,
title = {Design optimization of a pixel-based range telescope for proton computed tomography},
journal = {Physica Medica},
volume = {63},
pages = {87-97},
year = {2019},
issn = {1120-1797},
doi = {https://doi.org/10.1016/j.ejmp.2019.05.026},
author = {H. E. S. Pettersen and J. Alme and G. G. Barnaföldi and R. Barthel and A. {van den Brink} and M. Chaar and V. Eikeland and A. García-Santos and G. Genov and S. Grimstad and O. Gr{\o}ttvik and H. Helstrup and K. F. Hetland and S. Mehendale and I. Meric and O. H. Odland and G. Papp and T. Peitzmann and P. Piersimoni and A. {Ur Rehman} and M. Richter and A. T. Samn{\o}y and J. Seco and H. Shafiee and E. V. Skjæveland and J. R. S{\o}lie and G. Tambave and K. Ullaland and M. Varga-Kofarago and L. Volz and B. Wagner and S. Yang and D. Röhrich},
}

@ARTICLE{BergenpCTStatusReport,
AUTHOR={Alme, J. and Barnaföldi, G. G. and Barthel, R. and Borshchov, V. and Bodova, T. and {van den Brink}, A. and Brons, S. and Chaar, M. and Eikeland, V. and Feofilov, G. and Genov, G. and Grimstad, S. and Gr{\o}ttvik, O. and Helstrup, H. and Herland, A. and Hilde, A. E. and Igolkin, S. and Keidel, R. and Kobdaj, C. and van der Kolk, N. and Listratenko, O. and Malik, Q. W. and Mehendale, S. and Meric, I. and Nesb{\o}, S. V. and Odland, O. H. and Papp, G. and Peitzmann, T. and Pettersen, H. E. S. and Piersimoni, P. and Protsenko, M. and {Ur Rehman}, A. and Richter, M. and Röhrich, D. and Samn{\o}y, A. T. and Seco, J. and Setterdahl, L. and Shafiee, H. and Skjolddal, {\O}. J. and Solheim, E. and Songmoolnak, A. and Sudár, \'A. and S{\o}lie, J. R. and Tambave, G. and Tymchuk, I. and Ullaland, K. and Underdal, H. A. and Varga-Köfaragó, M. and Volz, L. and Wagner, B. and Wider{\o}e, F. M. and Xiao, R. and Yang, S. and Yokoyama, H.},   
TITLE={A High-Granularity Digital Tracking Calorimeter Optimized for Proton CT},      
JOURNAL={Frontiers in Physics},      
VOLUME={8},           
YEAR={2020},
DOI={10.3389/fphy.2020.568243},      
ISSN={2296-424X},
}

@article{bragg1904lxxiv,
  title={LXXIV. On the ionization curves of radium},
  author={Bragg, William Henry and Kleeman, RLXXIV},
  journal={The London, Edinburgh, and Dublin Philosophical Magazine and Journal of Science},
  volume={8},
  number={48},
  pages={726--738},
  year={1904},
  publisher={Taylor \& Francis}
}

@article{huda2007patient,
  title={Patient radiation doses from adult and pediatric CT},
  author={Huda, Walter and Vance, Awais},
  journal={American Journal of Roentgenology},
  volume={188},
  number={2},
  pages={540--546},
  year={2007},
  publisher={American Roentgen Ray Society}
}

@article{https://doi.org/10.1002/mp.15283,
author = {Bär, Esther and Volz, Lennart and Collins-Fekete, Charles-Antoine and Brons, Stephan and Runz, Armin and Schulte, Reinhard Wilhelm and Seco, Joao},
title = {Experimental comparison of photon versus particle computed tomography to predict tissue relative stopping powers},
journal = {Medical Physics},
volume = {49},
number = {1},
pages = {474-487},
doi = {https://doi.org/10.1002/mp.15283},
url = {https://aapm.onlinelibrary.wiley.com/doi/abs/10.1002/mp.15283},
eprint = {https://aapm.onlinelibrary.wiley.com/doi/pdf/10.1002/mp.15283},
year = {2022}
}

@article{hurley2012water,
  title={Water-equivalent path length calibration of a prototype proton CT scanner},
  author={Hurley, RF and Schulte, RW and Bashkirov, VA and Wroe, AJ and Ghebremedhin, A and Sadrozinski, HF-W and Rykalin, V and Coutrakon, G and Koss, P and Patyal, B},
  journal={Medical physics},
  volume={39},
  number={5},
  pages={2438--2446},
  year={2012},
  publisher={Wiley Online Library}
}

@article{Dickmann_2019,
doi = {10.1088/1361-6560/ab2474},
url = {https://dx.doi.org/10.1088/1361-6560/ab2474},
year = {2019},
month = {jul},
publisher = {IOP Publishing},
volume = {64},
number = {14},
pages = {145016},
author = {Dickmann, J and Wesp, P and Rädler, M and Rit, S and Pankuch, M and Johnson, R P and Bashkirov, V and Schulte, R W and Parodi, K and Landry, G and Dedes, G},
title = {Prediction of image noise contributions in proton computed tomography and comparison to measurements},
journal = {Physics in Medicine \& Biology}
}

@article{10.1002/mp.15644,
author = {Tattenberg, Sebastian and Madden, Thomas and Bortfeld, Thomas and Parodi, Katia and Verburg, Joost},
year = {2022},
month = {05},
pages = {},
title = {Range uncertainty reductions in proton therapy may lead to the feasibility of novel beam arrangements which improve organ‐at‐risk sparing},
volume = {49},
journal = {Medical Physics},
doi = {10.1002/mp.15644}
}

@article{10.1088/0031-9155/60/19/7585,
author = {Arbor, Nicolas and Dauvergne, Denis and Dedes, George and Létang, Jm and Parodi, K and Quiñones, Catherine and Testa, Etienne and Rit, Simon},
year = {2015},
month = {09},
pages = {7585-7599},
title = {Monte Carlo comparison of x-ray and proton CT for range calculations of proton therapy beams},
volume = {60},
journal = {Physics in medicine and biology},
doi = {10.1088/0031-9155/60/19/7585}
}

@article{byun2021physical,
  title={Physical and biological characteristics of particle therapy for oncologists},
  author={Byun, Hwa Kyung and Han, Min Cheol and Yang, Kyungmi and Kim, Jin Sung and Yoo, Gyu Sang and Koom, Woong Sub and Kim, Yong Bae},
  journal={Cancer Research and Treatment: Official Journal of Korean Cancer Association},
  volume={53},
  number={3},
  pages={611--620},
  year={2021},
  publisher={Korean Cancer Association}
}

@article{paganetti2012range,
  title={Range uncertainties in proton therapy and the role of Monte Carlo simulations},
  author={Paganetti, Harald},
  journal={Physics in medicine and biology},
  volume={57},
  number={11},
  pages={R99--R117},
  year={2012},
  publisher={IOP Publishing}
}

@article{bashkirov2016development,
  title={Development of proton computed tomography detectors for applications in hadron therapy},
  author={Bashkirov, Vladimir A and Johnson, Robert P and Sadrozinski, Hartmut F-W and Schulte, Reinhard W},
  journal={Nuclear Instruments and Methods in Physics Research Section A: Accelerators, Spectrometers, Detectors and Associated Equipment},
  volume={809},
  pages={120--129},
  year={2016},
  publisher={Elsevier}
}

@article{scaringella2023infn,
  title={The INFN proton computed tomography system for relative stopping power measurements: calibration and verification},
  author={Scaringella, Monica and Bruzzi, Mara and Farace, Paolo and Fogazzi, Elena and Righetto, Roberto and Rit, Simon and Tommasino, Francesco and Verroi, Enrico and Civinini, Carlo},
  journal={Physics in Medicine \& Biology},
  volume={68},
  number={15},
  pages={154001},
  year={2023},
  publisher={IOP Publishing}
}

@article{haberer2004heidelberg,
  title={The Heidelberg ion therapy center},
  author={Haberer, Th and Debus, J{\"u}rgen and Eickhoff, H and J{\"a}kel, Oliver and Schulz-Ertner, Daniela and Weber, Uli},
  journal={Radiotherapy and Oncology},
  volume={73},
  pages={S186--S190},
  year={2004},
  publisher={Elsevier}
}

@article{shepp1974fourier,
  title={The Fourier reconstruction of a head section},
  author={Shepp, Lawrence A and Logan, Benjamin F},
  journal={IEEE Transactions on nuclear science},
  volume={21},
  number={3},
  pages={21--43},
  year={1974},
  publisher={IEEE}
}

@article{sarrut2026gate,
  title={GATE 10 Monte Carlo particle transport simulation: I. Development and new features},
  author={Sarrut, David and Arbor, Nicolas and Baudier, Thomas and Bert, Julien and Chatzipapas, Konstantinos and Favaretto, Martina and Fuchs, Hermann and Grevillot, Loic and Harb, Hussein and Van Hoey, Gert and others},
  journal={Physics in Medicine \& Biology},
  volume={71},
  number={1},
  pages={015042},
  year={2026},
  publisher={IoP Publishing}
}

@article{article,
author = {Bimbot, R. and Geissel, H. and Paul, H. and Schinner, A. and Sigmund, Peter and Arista, Neyda and Mikkelsen, H. and Sørensen, A. and Thwaites, David},
year = {2005},
month = {06},
pages = {0-0},
title = {STOPPING OF IONS HEAVIER THAN HELIUM},
volume = {5},
journal = {Journal of the International Commission on Radiation Units and Measurements},
doi = {10.1093/jicru/ndi001}
}

@article{berger1992estar,
  title={ESTAR, PSTAR, and ASTAR: Computer programs for calculating stopping-power and range tables for electrons, protons, and helium ions},
  author={Berger, Martin J},
  journal={Unknown},
  year={1992}
}

@article{andersen1984simultaneous,
  title={Simultaneous algebraic reconstruction technique (SART): a superior implementation of the ART algorithm},
  author={Andersen, Anders H and Kak, Avinash C},
  journal={Ultrasonic imaging},
  volume={6},
  number={1},
  pages={81--94},
  year={1984},
  publisher={Elsevier}
}

@article{schulte2008maximum,
  title={A maximum likelihood proton path formalism for application in proton computed tomography},
  author={Schulte, RW and Penfold, SN and Tafas, JT and Schubert, KE},
  journal={Medical physics},
  volume={35},
  number={11},
  pages={4849--4856},
  year={2008},
  publisher={Wiley Online Library}
}

@article{krah2018comprehensive,
  title={A comprehensive theoretical comparison of proton imaging set-ups in terms of spatial resolution},
  author={Krah, Nils and Khellaf, Feriel and L{\'e}tang, Jean Michel and Rit, Simon and Rinaldi, Ilaria},
  journal={Physics in Medicine \& Biology},
  volume={63},
  number={13},
  pages={135013},
  year={2018},
  publisher={IOP Publishing}
}

@article{williams2004most,
  title={The most likely path of an energetic charged particle through a uniform medium},
  author={Williams, David C},
  journal={Physics in Medicine \& Biology},
  volume={49},
  number={13},
  pages={2899--2911},
  year={2004}
}

@article{polyak1990new,
  title={New stochastic approximation type procedures},
  author={Polyak, Boris T},
  journal={Automat. i Telemekh},
  volume={7},
  number={98-107},
  pages={2},
  year={1990}
}

@article{heder2022past,
  title={The past, present and future of the ELKH cloud},
  author={H{\'e}der, Mih{\'a}ly and Rig{\'o}, Ern{\H{o}} and Medgyesi, Dorottya and Lovas, R{\'o}bert and Tenczer, Szabolcs and Farkas, Attila and Em{\H{o}}di, M{\'a}rk Benj{\'a}min and Kadlecsik, J{\'o}zsef and Kacsuk, P{\'e}ter},
  journal={Inform{\'a}ci{\'o}s T{\'a}rsadalom: T{\'a}rsadalomtudom{\'a}nyi Foly{\'o}irat},
  volume={22},
  number={2},
  pages={128--137},
  year={2022}
}

\end{document}